\documentclass[epj]{svjour}
\usepackage{atbegshi}
\AtBeginDocument{\AtBeginShipoutNext{\AtBeginShipoutDiscard}}

\usepackage{amsmath}	
\usepackage{color}
\usepackage{mathtools, cuted}
\usepackage{hyperref}
\pdfoutput=1
\newcommand{\Liel}{$^{11}$Li}
\newcommand{\Lin}{$^{9}$Li}

\begin{document}
	\title{Characterization of vorticity in pygmy resonances and soft-dipole modes with two-nucleon transfer reactions}
	\author{R.A. Broglia\inst{1,}\inst{2} \and F. Barranco\inst{3} \and G. Potel\inst{4} \and E. Vigezzi\inst{5}}
	\institute{The Niels Bohr Institute, University of Copenhagen, 
		DK-2100 Copenhagen, Blegdamsvej 17, Denmark  \and Dipartimento di Fisica, Universit\`a degli Studi Milano,
		Via Celoria 16, 
		I-20133 Milano, Italy \and Departamento de F\`isica Aplicada III,
		Escuela Superior de Ingenieros, Universidad de Sevilla, Camino de los Descubrimientos, 	Sevilla, Spain
		\and National Superconducting Cyclotron Laboratory, Michigan State University, East Lansing, Michigan 48824, USA
		\and INFN Sezione di  Milano,
		Via Celoria 16,  I-20133 Milano, Italy }
	\begin{abstract}
		{The properties of the two-quasiparticle-like soft E1-modes and PDR have been and are 
			systematically studied with the help of inelastic and electromagnetic experiments which essentially probe the particle-hole 
			components of these vibrations. It is shown that further insight in their characterisation 
			can be achieved with the help of two-nucleon transfer
			reactions, in particular concerning the particle-particle components of the modes, in terms of absolute 
			differential cross sections which take properly into account successive and simultaneous transfer mechanisms corrected for non-orthogonality,
			able to reproduce the experimental  findings at the 10\% level. The process $^9$Li$(t,p)^{11}$Li(1$^-$) is
			discussed, and absolute cross sections predicted.} 	
	\end{abstract}
	\PACS{21.60.Jz,23.40.-s,26.30.-k}
\titlerunning{Pygmy resonances and two--nucleon transfer reactions}
\maketitle
\section{Background for subject and title}
\label{intro}
Almost two decades ago the paper entitled ``The halo of the exotic nucleus $^{11}$Li: a single Cooper pair'' we 
(FB,RAB,
EV and G. Col\`o) wrote 
 with Pier Francesco Bortignon was published (Eur. Phys. J. {\bf A11}, 385-392 (2001)). Since we  started, few years before, thinking
 on this unstable, exotic nucleus, we became  mesmerized by the possibilities the system offered, as a femtometer  many-body laboratory to learn about 
 the origin of pairing in nuclei. In particular, in connection with the E1-soft mode acting as the tailored glue of the two halo neutrons.
 \footnote{The paper had been completed a year before, passed the Editorial desk of Nature, and was rejected by the referee, himself a major
 figure in the field of nuclear structure, as I learned when he walked to me and referred, after my (RAB) presentation at the Conference 
 ``Bologna 2000- Structure of Nuclei at the Dawn of the Century'', few months later, that the reason for the rejection had been the misunderstanding 
 concerning our use or less of the bare NN-$^1S_0$ pairing interaction. Interaction which is, in $^{11}$Li, subcritical to bind the two halo neutrons to the core $^9$Li.
An outcome which  Pier Francesco, who had dedicated much time and effort to peer review activity, being himself a superb practitioner of this difficult but 
necessary ``art'' could accept but not understand. He felt, in a deep sense, responsible of bringing the ``truth'' out of each paper he accepted to review.}
 Also to test the flexibility 
 of nuclear field theory (NFT)  to treat on equal footing both (strongly and weakly) bound as well as continuum states. Furthermore to assess NFT abilities 
 to lead to convergent results when simultaneously confronted with weak (induced interaction) and strong (self-energy and parity inversion) particle-vibration coupling 
 vertices \footnote{A main contribution of Pier Francesco's lifelong work in NFT (see e.g. \cite{Bortignon:78,Bortignon:77b})}. Our infatuation with the halo nuclei ``laboratory'' did not falter with the years, and the last paper we (the authors plus A. Idini) published with Pier Francesco
 on the subject, this time with the title ``Unified description of structure and reactions: implementing the nuclear field theory'', Phys. Scr. {\bf 91} 063012 (2016), presented an
 extension of NFT to simultaneously deal with structure and reactions, but also to spell in detail renormalization. 
 Pier Francesco's efforts to connect these developments to well established results from   the literature making use of his profound insight concerning the concepts and techniques of many-body
 physics, constituted a major inspiration. 
 Pier Francesco's voice still rings in my (RAB) 
ears when he, after the task was completed, enthusiastically epitomised as ``forte'' the sentence introducing  the corresponding discussion, namely ``Enter empirical renormalization''.
Within this context it is only natural to dedicate the present paper to honour the memory of Pier Francesco, colleague and friend, reference point in finding the right path in our research.

\section{Introduction} 

The nature of the (dipole) low-energy strength (LES) can be theoretically  characterised by the response function (RF), the transition densities (TD) 
 and the transition currents 
(TC). While $dB(E1)/dE$ (e$^2$fm$^2$/MeV) response function can be observed  in e.g. ($\gamma,\gamma')$ experiments, the situation is less clear concerning
TD and TC \cite{Repko:13,Ryezayeva:02}.

Experimentally, one has observed correlation existing  between inelastic scattering reactions  (ISR) and two-particle transfer reactions (TPTR) \cite{Broglia:71}. The fact that 
ground state correlations  (GSC) associated with a particle-hole ($ph$) collective vibrations contribute constructively (des\-tructive\-ly) 
coherent to the absolute ISR (TPTR) differential 
cross section while the opposite is true in the case of a ($pp$) and ($hh$) collective vibrations, can help at shedding light into the possible relations between the observable $B(E1)$ response function,
and the associated theoretical TD and TC of dipole LES. In particular in connection with the soft $E1$-mode of $^{11}$Li. The reasons for this selection are discussed in  the
next section. 

\section{Nuclear embodiment of Cooper model and of a vortex}

\begin{figure*}
\centerline{\includegraphics[width=15cm]{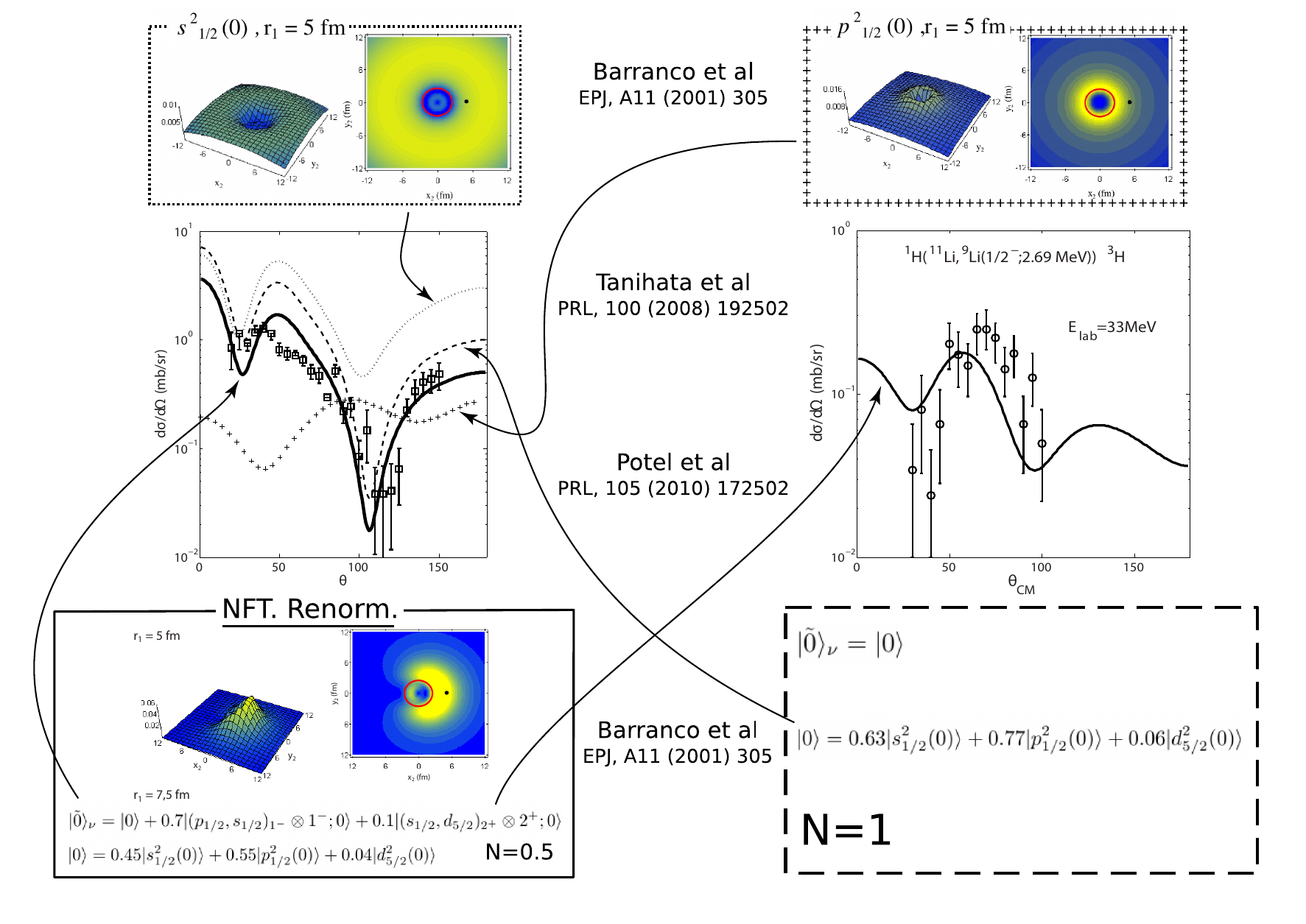}}
\caption{Absolute, two-nucleon transfer differential cross section associated with  the ground
state and the first excited state of $^9$Li, excited in the reaction $^1$H($^{11}$Li,$^9$Li)$^3$H in comparison
with the predicted differential cross sections worked out making use of spectroscopic amplitudes and Cooper pair wavefunctions
calculated with nuclear field theory.}
\label{fig:1}       
\end{figure*}

Because  the $N=6$ closed shell isotope $^{9}_3$Li$_6$ is a well bound system and $^{10}_3$Li$_7$ with one neutron outside closed shell is not,
while $^{11}$Li with two neutrons outside closed shells  is bound, 
one can posit pairing to be at the basis of the binding of the two halo neutrons to the core $^9$Li. In keeping with the fact that 
{\bf a)} the two-neutron separation energy of the latter system is $S_{2n}$= 6.094 MeV, while that of $^{11}$Li is $S_{2n}$= 0.380 MeV and,
{\bf b)} that  the corresponding radii are $R= $2.7 fm and $R= 4.6$ fm, one can view the ground state of \Liel, considering the 
$p_{3/2}(\pi)$ odd proton as a spectator, to be the nuclear embodiment of a  Cooper pair:   two fermions moving in time 
reversal states 
(coupled to  $J^{\pi} = 0^+$), weakly interacting  on top of a
quiescent Fermi sea, leading to a very extended (halo), barely bound (quasi boson) system (\cite{Cooper:56}, see also \cite{Barranco:01})\footnote{Removing one of the fermions the other becomes also unbound. Such a system
is, in nuclei, an elementary mode of excitation namely a pair addition mode which in the case of $^{11}$Li becomes
a (halo) pair addition mode. Mode which is specifically probed by two-neutron (Cooper pair) transfer (tunnelling). Mode which can,
in principle, be moved around and, arguably, found as the first excited $0^+$ halo state of $^{12}$Be. That one dubs such pairing
vibrational mode with transfer quantum number $\beta=+2$ a Borromean entity (see e.g. \cite{Zhukov:93} and refs. therein), gives  only
one aspect  of the physics  at the basis of the BCS explanation of superconductivity, paradigm of spontaneous symmetry
breaking theories, and of the prediction of the Josephson effect \cite{Josephson:62} which has provided  relative voltage standards of 1 part in
$10^{19}$.} (see Fig. \ref{fig:1}, also
Appendix A) .
Specific, detailed support for  this picture is provided by 
the two-neutron transfer process $^1$H(\Liel,$^9$Li(gs))$^3$H. Theory leads to  a quantitative account of the experimental findings at the 10\% level \cite{Tanihata:08,Potel:10}.

Within this scenario the soft E1-mode (again viewing the odd $p_{3/2}(\pi)$ proton  as a spectator), observed at $E_x \leq 1$ MeV with a width of $\Gamma \approx 0.5$ MeV
and carrying  $\approx 6-8$\% of the TRK energy weighted sum rule \cite{Kanungo:15,Nakamura:06}, can be viewed as a $J^{\pi}= 1^-$ Cooper pair, that is a quantal nuclear vortex \cite{Avogadro:07,Avogadro:08},
namely  the quantal structure which allows rotation around a symmetry axis. 

The calculation of this soft E1 mode was carried out making use of Saxon-Woods potentials  to approximately reproduce 
both protons and neutron  single-particle states, and BCS to determine the associated occupation numbers. The continuum was discretised  by placing the system in a spherical box of radius $R_{box}=$ 40 fm, ensuring convergence. 
The QRPA solution of the full dipole response of $^{11}$Li was determined making use of a separable dipole-dipole interaction of self-consistent strength in a two-quasiparticle basis 
with energies up to 50 MeV. The results lead to a soft mode of centroid 0.75 MeV and FWHM $\Gamma=$ 0.5 MeV, and carrying 6.2 \% of the EWSR, in overall
agreement  with the experimental findings. The associated $r^2-$weighted transition densities are typical of a dipole LES of light halo systems: 
out-of-phase motion of the neutron (skin) halo against a core in which protons and neutrons oscillate in phase \cite{Broglia:19} (Fig. 2) .

The calculations discussed above are adequate  to describe the processes associated with the role of intermediate  boson played
by the two halo neutron to the core $^9$Li (induced pairing interaction). 
Also to provide the basis to calculate the absolute differential cross sections and transition probabilities associated with the probing 
of the mode with inelastic scattering and particle transfer experiments (see Figs. 3 and 4 below).

However, if one would like to investigate the quantitative consequences of the interplay between particle-hole ($ph$)- and
particle-particle ($pp$)-like ground state correlations (GSC) one is forced to go beyond QRPA, and take into account the variety of renormalisation processes. Namely, self-energy of single-particle states and vertex corrections (see e.g. 
\cite{Barranco:01,Lenske:01,Orrigo:09,Barranco:19,Broglia:16}). An ambitious, pluriannual project going beyond the framework of the present 
contribution. 

Concerning the interplay between ($pp$)- and ($ph$)-GSC, although  seemingly new,  this subject, namely the theoretical discussion and the experimental consequences of $pp$ and $ph$ contributions to nuclear vibrations has a long tradition.
In particular in connection with $\beta-$ and pairing-vibrations in superfluid quadrupole deformed nuclei, 
where they get mixed due to the spontaneous breaking of rotational invariance \footnote{Similar to what happens in $^{11}$Li  between the dipole pair addition mode  and the soft
E1 mode, associated with a system poised to acquire a permanent dipole deformation (see App. B).}. A subject started  in the 1960's \cite{Bes:63} and still very much open  (see \cite{Aprahamian:18,Sharpey:19}
and refs. therein). One knows of only few cases  in which, due to a propitious distribution of single-particle levels around the Fermi energy, one can observe 
a  clear signal, for example an important enhancement  of the two-nucleon transfer absolute cross section typical of pairing vibrations around closed shell 
systems rather than of superfluid nuclei (see \cite{Maher:70,Casten:72,Ragnarsson:76} and refs. therein). 
Arguably, $^{11}$Li  constitutes such a propitious case concerning the characterisation of the dipole LES.

Let us close this section by reminding  that the appearance of quantised vortices constitutes a hallmark of superfluidity.
In a superfluid, a quantum vortex carries quantised orbital angular momentum, being 
zeros of the wave function around which the velocity field has a solenoidal shape. 
A nucleus acting as impurity immersed in a Wigner-Seitz cell of the inner crust of a neutron star 
(roughly equivalent to $^{1000}_{50}$Sn) experiences that a vortex becomes pinned by skating around it
along the nuclear surface \cite{Avogadro:07,Avogadro:08}. This is in keeping with the fact that the sequence of levels 
of nuclei along the stability valley display, around the Fermi energy, a distribution of single-particle levels all
carrying, exception made for the intruder one, the same parity. In the case of the neutron drip line 
nucleus $^{11}$Li, the outermost neutrons move  with essentially equal amplitude ($\approx 0.7$) in the 
almost degenerate $s_{1/2}$ and $p_{1/2}$ halo orbitals. Coupling the halo neutrons to angular momentum and parity 
$1^-$ leads to a quantal vortex (Cooper pair with $J^{\pi}=1^-$) which again skates on the neutron (halo) skin.
Said it differently, the soft $E1$-mode of $^{11}$Li can be viewed as a example of a quantum vortex in a nucleus 
\footnote{This result provides also an answer to Nambu's last question in Section 
{\it Conclusions and speculations}  of \cite{Mukerjee:89}.}.

\begin{figure*}[h!]
\centerline{\includegraphics[width=8cm]{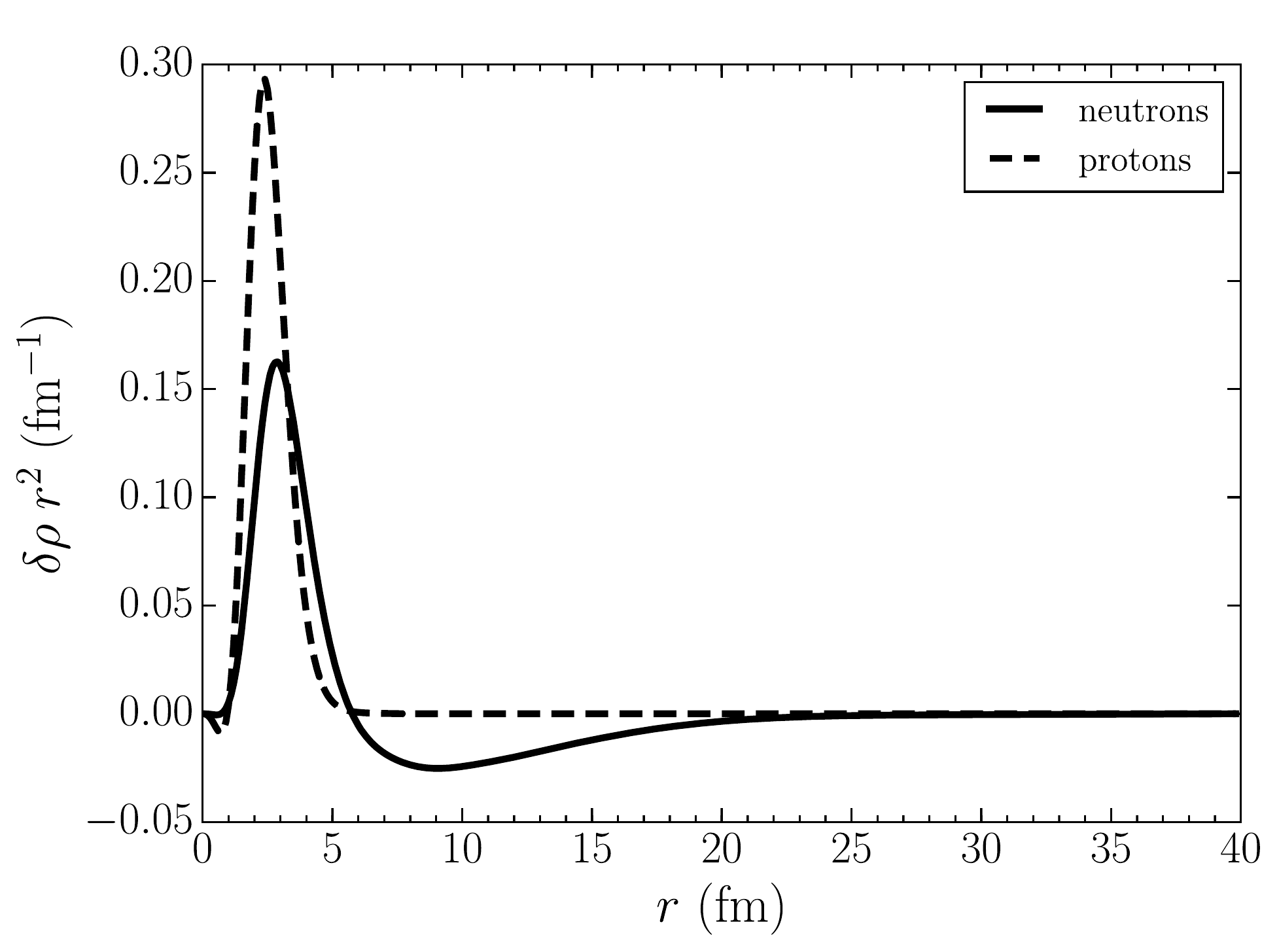}}
\caption{Transition density multiplied by $r^2$ associated with states representative of the soft dipole mode of $^{11}$Li .}
\label{fig:2}      
\end{figure*}

Let us now think in terms of a quantum superfluid.  The ground state of $^{11}$Li 
can be expressed as $|gs(^{11}$Li)$\rangle  = |\Psi \otimes 1p_{3/2}(\pi)\rangle $, where 
$\Psi= \sqrt{\alpha'_0} e^{i\phi}$ parallels the Ginzburg-Landau wave function, $\phi$ the gauge angle 
and $|\Psi|^2 = \alpha'_0$ the superfluid density. In the present case it coincides with the modulus 
square of the halo neutron Cooper pair wavefunction \footnote{In other words, we are taking the first term of 
Eq. (A.1), namely the Cooper pair component expressed in terms of renormalised states
($\tilde j^2(0))$ and amplitudes (note that the state in Eq. (A.2) is normalised to $\approx 1/2)$. One could argue that a superfluid 
can hardly be made out of a single Cooper pair. This may be true but for this purpose  one Cooper pair is hardly much worse than 
 five  Cooper pairs, the case of $^{120}$Sn, paradigm of superfluid nuclei. In any case, one can correctly view
the $^{11}$Li ground state as a pairing vibrational mode, vibrations which are very collective in nuclei, describe it in the RPA
and construct with the corresponding $X$ and $Y$ amplitudes, the effective BCS $U_{eff}$ and $V_{eff}$ factors 
\cite{Potel:13b}.}. Within this scenario the soft $E1$-mode can be described in terms of the QRPA.
It constitutes a two-quasiparticle, large amplitude mode, the estimated number of crossings being\footnote{$n \approx (1/4\sqrt{\pi})\times A \times \beta$(\cite{Barranco:88,Barranco:90} see also \cite{Brink:05} Eq. (7.35)), with 
	$\beta \approx r_{extr}/R(^{11}Li) \approx 1.7$, the radius of $^{11}$Li being $R(^{11}Li$) = 4.6 fm, while
	the extreme of the $r^2-$weighted halo transition density takes place at $r_{extr} \approx $ 8 fm (Fig. \ref{fig:2}).} $n \approx 3$.
 We are thus confronted with a two-quasiparticle  mode of  angular momentum and parity $1^-$. Again the scenario of a quantal nuclear vortex.

\begin{figure*}[h!]
\centerline{\includegraphics[width=15cm]{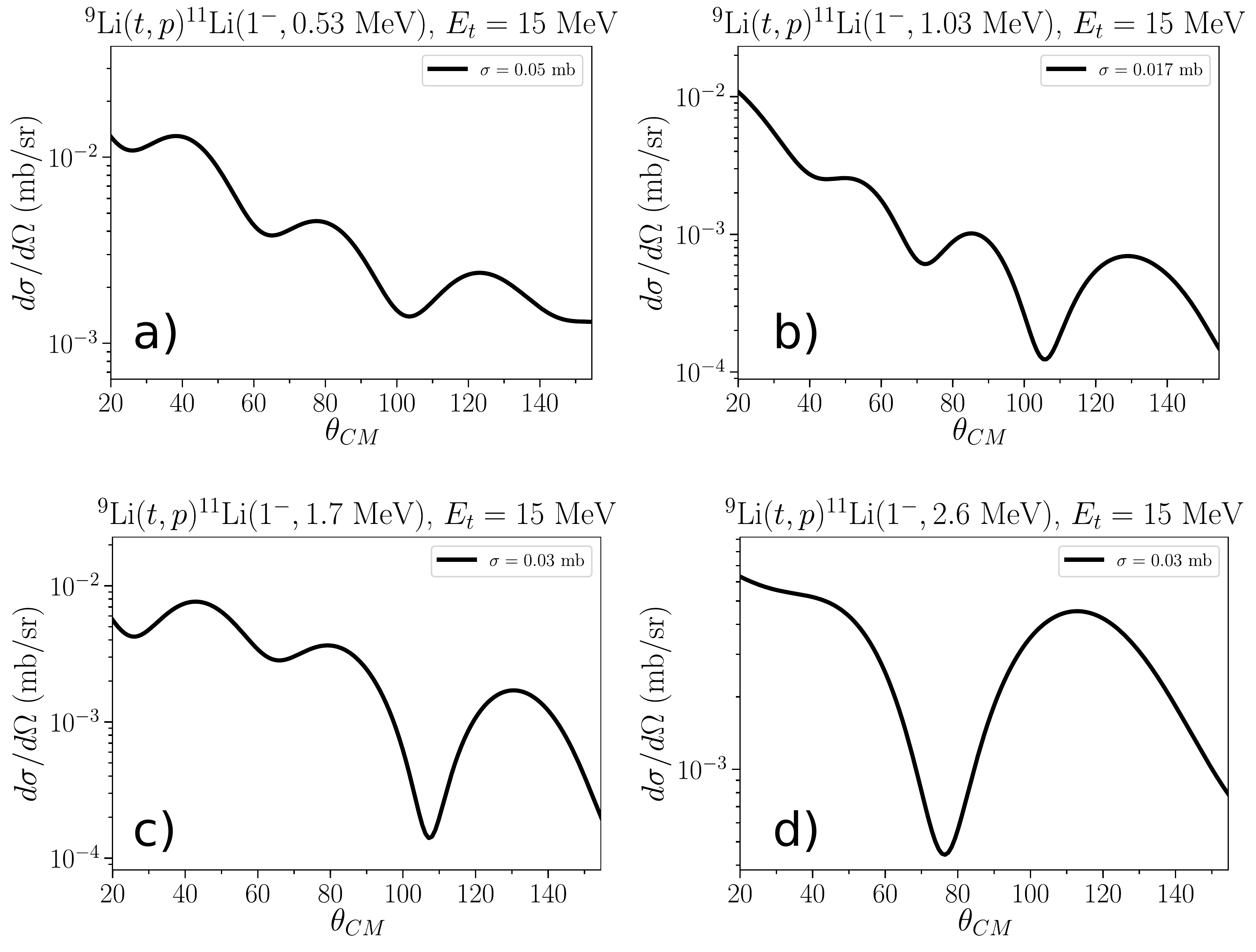}}
\caption{Absolute  two-nucleon differential cross section associated with 
 states  representative of the soft dipole mode of $^{11}$Li. The total 
absolute cross section results from the integration of $d\sigma/d\Omega$ in the angular 
interval (20$^\circ \leq \theta_{CM} \leq 154.5^{\circ}$ \cite{Tanihata:08}).}
\label{fig:3}       
\end{figure*}

\section{Two-nucleon transfer and detailed balance}

Our aim is that of calculating the absolute differential cross section associated with the reaction 
\begin{equation}
^{9}Li + t \to \; ^{11}Li (1^-) + p
\end{equation}
where the label 1$^-$ indicates states belonging to the soft E1-mode.  But before doing so, we will proceed at assessing the accuracy with 
which such calculations can be carried out.  This is in keeping with the fact that the prediction of (t,p) differential cross sections 
and the relative contribution to them of GSC
can be of use to compare at profit  with the results of active target and inverse kinematics experiments, to the extent that absolute cross sections can be worked out with sensible small errors, well below the 30\% level. 
The experimental and theoretical absolute integrated cross sections associated with $^{11}$Li$(p,t)$$^9$Li(gs) for a bombarding energy of 3.3 MeV/A (Fig. 1), are 
$\sigma_{exp}(gs)= 5.7 \pm$ 0.9 mb \cite{Tanihata:08}  and $\sigma_{th} (gs)$ = 6.1 mb \cite{Potel:10} respectively, implying a deviation of the order of 7\% (see also \cite{Potel:13}). 
Let us now connect the (p,t) to (t,p) reaction through detailed balance, namely 
\begin{equation}
g_{\alpha} k_{\alpha}^2 \frac{d\sigma}{d\Omega} (\alpha \to \beta) = g_{\beta}k_{\beta}^2 \frac{d\sigma}{d \Omega}(\beta \to \alpha)
\end{equation}
where \footnote{Concerning the notation see \cite{Broglia:04a}.} $\alpha=(^9$Li(gs)+t)  and  
$\beta=(^{11}$Li(gs)+p), $k_{\alpha},k_{\beta}$ and $g_{\alpha},g_{\beta}$ being the 
relative linear momentum and the total number of spins in entrance and exit channel respectively. In the case under discussion
$g_{\alpha}=g_{\beta}= 3/2$ while $k_{\alpha}^2 =$ 7.425 fm$^{-2}$ and 
$k_{\beta}^2$= 0.72 fm$^{-2}$ resulting in $\sigma(\alpha \to \beta)$= 0.097$\times\sigma(\beta \to \alpha)$ = 0.55 $\pm 0.09 $ mb.
As expected, the  microscopic calculations carried out making use of the elements --wave functions and thus two-nucleon spectroscopic amplitudes, single-particle wave functions and 
optical potentials--  and of the two-particle-transfer code \textsc{cooper} \cite{Potel:x} 
employed in \cite{Potel:10}  leads  to the detailed balance result. 

\begin{figure*}[h!]
\centerline{\includegraphics[width=20cm]{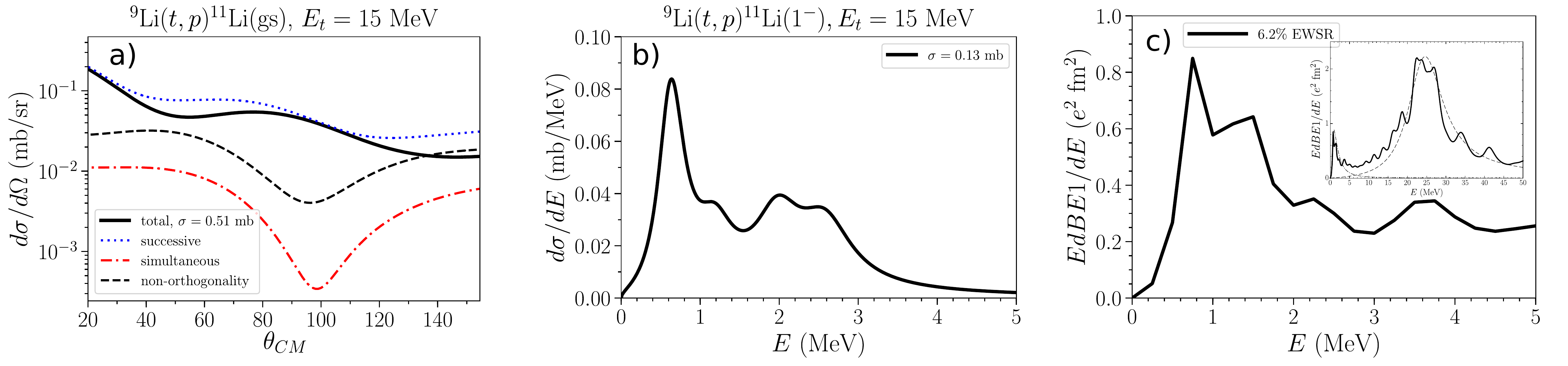}}
\caption{(a) Absolute differential cross section associated with the ground state 
transition. The total absolute cross sections are calculated as explained in the caption to Fig. \ref{fig:3}; (b) 
two-neutron transfer cross dipole strength function $d \sigma(1^-)/dE$, (c) EWSR $E dB\,E1/dE$ as
a function of the energy.}
\label{fig:4}       
\end{figure*}
 
 In what follows we calculate, for a triton bombarding energy of 15 MeV, the absolute 
 differential cross section associated with  the reaction (1) for the states associated with the soft $E1$-mode,
 examples of which are shown inf Fig. 3, as well as for the ground state  transition 
 $^9$Li$(t,p)^{11}$Li(gs) (Fig. \ref{fig:4}(a)). Broadening each individual dipole state with a Lorentzian function 
 and making use of the corresponding integrated cross section  within the same angular range
 as that of the ground state transition ($20^{\circ} \leq \theta_{CM} \leq 154.5^{\circ}$ \cite{Tanihata:08}), the 
 two-neutron transfer dipole strength function $d\sigma(1^-)/dE$ (mb/MeV) was constructed, and is displayed 
 in Fig. 4(b). The subtended area (energy integration) is $\sigma(1^-) = 0.13 $ mb, to be compared with $\sigma(gs)= $ 0.51 mb. 
 In all cases, the successive transfer of the two neutrons dominates over the simultaneous transfer (Fig. \ref{fig:4}(a)). The reason
 being the very poor overlap between the very extended single-particle neutron wave function involved in the structure of the soft E1-mode 
 and that associated with the triton. Within this context, we are technically in a two-center shell mode
 situation. 
 
Similarly to $d\sigma(1^-)/dE$, making use of the transition densities associated with the dipole states, the dipole strength function
$dB(E1)/dE$ and associated Thomas-Reiche-Kuhn sum rule ($E dB(E1)/dE)$ was calculated as a function
of the energy. It is displayed in Fig. \ref{fig:4}(c). The subtended area (energy integration) leads to 6.2\% of the EWSR.

The similitude between the results displayed in Figs. 4(b) and 4(c) is apparent. 
The connection between theory and experiment (predictions which can be experimentally tested) are the  two-nucleon transfer form factors 
in the first case, and the particle-hole transition densities in the second case.  Ground state correlations 
affect quite differently (opposite) these theoretical quantities depending on whether they are 
of ($ph$) or ($pp$) type. Of particular interest are expected to be situations in which levels associated with the high 
(low) energy tail of the PDR (GDR) are close by in energy, so as to be able to probe on equal footing 
and with both TNTR and ISR large amplitude isoscalar modes with substantial  $pp$ components, with small amplitudes 
isovector vibrations of mainly $ph$ character.

 Summing up, we are in presence  of a collective mode peaked at $E_x \leq$ 1 MeV, displaying a transition density consistent 
 with that of a PDR, carrying a non-negligible fraction of the EWSR and an absolute two-neutron transfer cross section of the order of 25\% of the ground state one. 
 A laboratory to test, once the full renormalized nuclear field theory structure results become available, 
 the texture of the associated ground state correlations through inelastic and electromagnetic processes, as well as two-particle 
 transfer ones, processes which are specific to shed light in the ($ph$) and ($pp$) aspects of the correlations respectively.

 \section{Conclusions}
 
 The combination  of inelastic processes   and of two-particle transfer reactions  (strength functions) can be used at profit in characterizing  the dipole LES 
 provided  one is able to calculate (TNTR) absolute differential cross sections  at the 10\% level accuracy,  so as to be able to assess the role  played by ground
 state correlations within experimental error.  Such requirement may also imply, as in the present  case, the calculation  of absolute differential cross sections  to 
 continuum states.
 
 A number of  problems remain open, in particular concerning the fact that  the wavelength of $(\gamma,\gamma')$ 
 exciting  the dipole LES is much larger than nuclear dimensions, let alone the fact that two-particle transfer  is dominated by successive transfer, and that 
 the associated form factor receives  important contributions from configurations in which the two neutrons are  essentially a nuclear diameter apart. 
 This, together with the presence of substantial neutron excess leading to conspicuous isospin mixing,  sets a question mark on the ``observability''
 of clear cut, distinct  transition densities  and of velocity fields.  To which extent the measurement  of magnetic moments  could help at shedding light  on some of these
 questions, remains both an experimental and theoretical open problem useful to look at.

\appendix
\section{Analytic calculation of $^{11}$Li(p,t)\Lin(gs)}
\setcounter{equation}{0}
\makeatletter
\renewcommand{\theequation}{A\@arabic\c@equation}
\makeatother

 The $^{11}$Li  ground state wave function can be written, assuming the odd proton to act as
 a spectator,  as\\
 $|gs(^{11}$Li)$\rangle $= $|p_{3/2}(\pi)\rangle  |\tilde 0\rangle _{\nu}$ where,
 \begin{equation}
 |\tilde 0\rangle _{\nu} = |0\rangle  + |ind\rangle ,
 \end{equation}
 \begin{equation}
 |0\rangle _{\nu} = 0.45 |{\widetilde s_{1/2}}^2(0)\rangle  + 0.55 |{\widetilde p_{1/2}^2(0)}\rangle  + 0.04 |{\widetilde d_{5/2}^2(0)}\rangle ,
 \end{equation}
 and
 \begin{equation}
 |ind\rangle = 0.7 |({\widetilde s_{1/2}}, {\widetilde p_{1/2}})_{1^-}; 0^+\rangle  + 
             0.1 |({\widetilde s_{1/2}}, {\widetilde d_{5/2}})_{2^+}; 0^+\rangle ,
 \end{equation}
${\widetilde s_{1/2}}$, ${\widetilde p_{1/2}}$  and ${\widetilde d_{5/2}}$ being renormalised single-particle states
 of $^{10}$Li the first two lying at threshold  with energies ${\widetilde \epsilon_{\widetilde s_{1/2}}}$ = 0.2 MeV,
 ${\widetilde \epsilon_{\widetilde p_{1/2}}}$ = 0.5 MeV \cite{Barranco:19}.
 Making use of the associated renormalised single-particle wave function \cite{Barranco:01}, the two-nucleon transfer formfactor 
 associated with the reaction $^1$H($^{11}$Li,$^9$Li(gs))$^3$H
 was worked out  and 
 the absolute differential cross section calculated \cite{Potel:10}. It provides a quantitative account of the absolute 
 differential cross section and of the integrated experimental
 one \cite{Tanihata:08}: $\sigma_{exp}(gs)= 5.7 \pm 0.9$ mb, $\sigma_{th}(gs)= 6.1$ mb.
 
 Let us now use the essentially two-component wave function (A.2) (see Fig. 5), to calculate 
 this cross section in a simple fashion. 
 Making use of Woods-Saxon single-particle  wavefunctions as formfactors the absolute differential cross section of the pure configurations  $s^2_{1/2}(0)$ and $p^2_{1/2}(0)$
 were calculated. The integrated values are 
 \begin{equation}
  \sigma(s^2_{1/2}(0))= 23.17\; {\rm mb},
 \end{equation}
 \begin{equation}
 \sigma(p^2_{1/2}(0))= 2.16 \; {\rm mb}.
 \end{equation}
 Combining Eqs. (A.2)-(A.5) one  can estimate,
 \begin{equation}
 \sigma(gs)= (0.45 \sqrt{23.17} + 0.55 \sqrt{2.16})^2  \; {\rm mb} = 8.86\; {\rm mb} ,
 \end{equation}
 an estimate 45\% $\left( (8.86-6.1)/6.1 \right)$ in error  in relation to $\sigma_{th}(gs)$.
 
 At the basis of this discrepancy one finds the fact that  the values of the cross sections reported in Eqs. (A.4) and (A.5) were calculated  with bare and 
 not renormalised formfactors. A simple estimate of the modifications of these formfactors due to renormalisation and 
 associated couplings is related to the fact that $s_{1/2}$ experiences no centrifugal barrier as compared to the 
 $p_{1/2}$ wavefunction. Because concentrated single-particle wavefunctions lead to larger matrix elements, and
 thus to cross sections,
 than less localised ones (think of the difference between neutron and proton pairing matrix elements), the use of 
 renormalised formfactors  will reduce the cross section displayed in Eq. (A.4) with respect to that shown in Eq. (A.5). A
  simple estimate can be made in terms of an effective
 diffusivity (for \Liel),
 \begin{equation}
 a_{eff} = \frac{R(^{11}{\rm Li})} {R_0(^{11}{\rm Li})} \times a = 
 \frac{4.6}{2.7} \times 0.65 {\rm fm} \approx 1.1 {\rm fm},
  \end{equation}
 where R(\Liel) = 4.58 $\pm$ 0.13 fm is the measured radius of \Liel, while $R_0$(\Liel) = 1.2 $\times (11)^{1/3}$ fm = 2.7 fm.
 Assuming the $p_{1/2}$ single-particle wave function feels an effective potential of radius $R_0$(\Liel) + $a_{eff}$ = 3.8 fm,
  while the $s_{1/2}$ experience that of radius $R_0$(\Liel)+ 2 $ \times a_{eff}$ = 4.9 fm, one can estimate the relative decrease of the cross
  section (A.4) with respect to (A.5), i.e.: $ \left(\frac{3.8}{4.9}\right)^2 \times 23.17$ mb $\approx$ 14 mb.
  The simple estimate (A.6) 
 leads, in this case, to 
 \begin{equation}
 \sigma(gs)= (0.45 \sqrt{14}+ 0.55 \sqrt{2.16})^2 {\rm mb} \approx  6.2 {\rm mb},
 \end{equation}
 a quantity which deviates by $\approx 1\%$ ((6.2-6.1)/6.1) from $\sigma_{th}(gs)$. 
 In other words, a proper estimate of the cross sections shown in Eqs. (A.4) and (A.5)
 should be carried out making use of the renormalised radial wave functions
  ${\widetilde R_{s_{1/2}}}$ and ${\widetilde R_{p_{1/2}}}$.
 
 The large error to be associated with the analysis of the experimental cross sections in terms of simple 
 theoretical estimates making use of Woods-Saxon formfactors is rather suggestive. While  one would not attempt a quantitative analysis of the spectrum of
 nuclei in terms of Woods-Saxon single-particle  energies such an approach is still common praxis in attempting at extracting 
 spectroscopic factors from transfer data. 
  The deep interweaving existing between structure and reactions implies that structure amplitudes and  reaction formfactors should be calculated at
 the same level of accuracy (renormalization)
  to be able to compare at profit theory with experiments (see e.g. \cite{Broglia:16}).

 \section{\\(Spontaneously broken) spin-orbit symmetry, tensor force}
\setcounter{equation}{0}
\makeatletter
\renewcommand{\theequation}{B\@arabic\c@equation}
\makeatother

 The essence of BCS pairing  mechanism is Cooper pair  formation due to a phonon-induced attraction \cite{Bardeen:57a,Bardeen:57b}. It leads to a non zero pair correlation function, 
 $\langle \sum_{\nu} a^{\dagger}_{\nu}a^{\dagger}_{\tilde \nu} \rangle  = \alpha'_0 e^{-2i \phi}$, and thus 
 to dynamical (spontaneous) symmetry breaking in gauge space. The magnitude of the associated deformation 
 is measured by the number of Cooper  pairs $\alpha'_0$ which, divided by the appropriate volume leads to the pair density. It defines a privileged orientation in gauge space subtending a gauge angle $\phi$ 
 with respect to the laboratory system of reference.

The chain of events ``crystal formation'' $\to$ ``superconductivity'' and associated violation of translational and gauge invariance, 
provides an example of the theoretical possibilities, tumbling in this case, related with spontaneously broken symmetry, 
in keeping with the fact that phonons are the Goldstone modes restoring Galilean invariance. 
In contrast, or more correctly, in parallel to the tumbling chain of symmetry breakings of descending energy scales, the idea of bootstrap
is that the chain is circular and self sustaining \cite{Nambu:91}.
\begin{figure*}
	\centerline{\includegraphics[width=18cm]{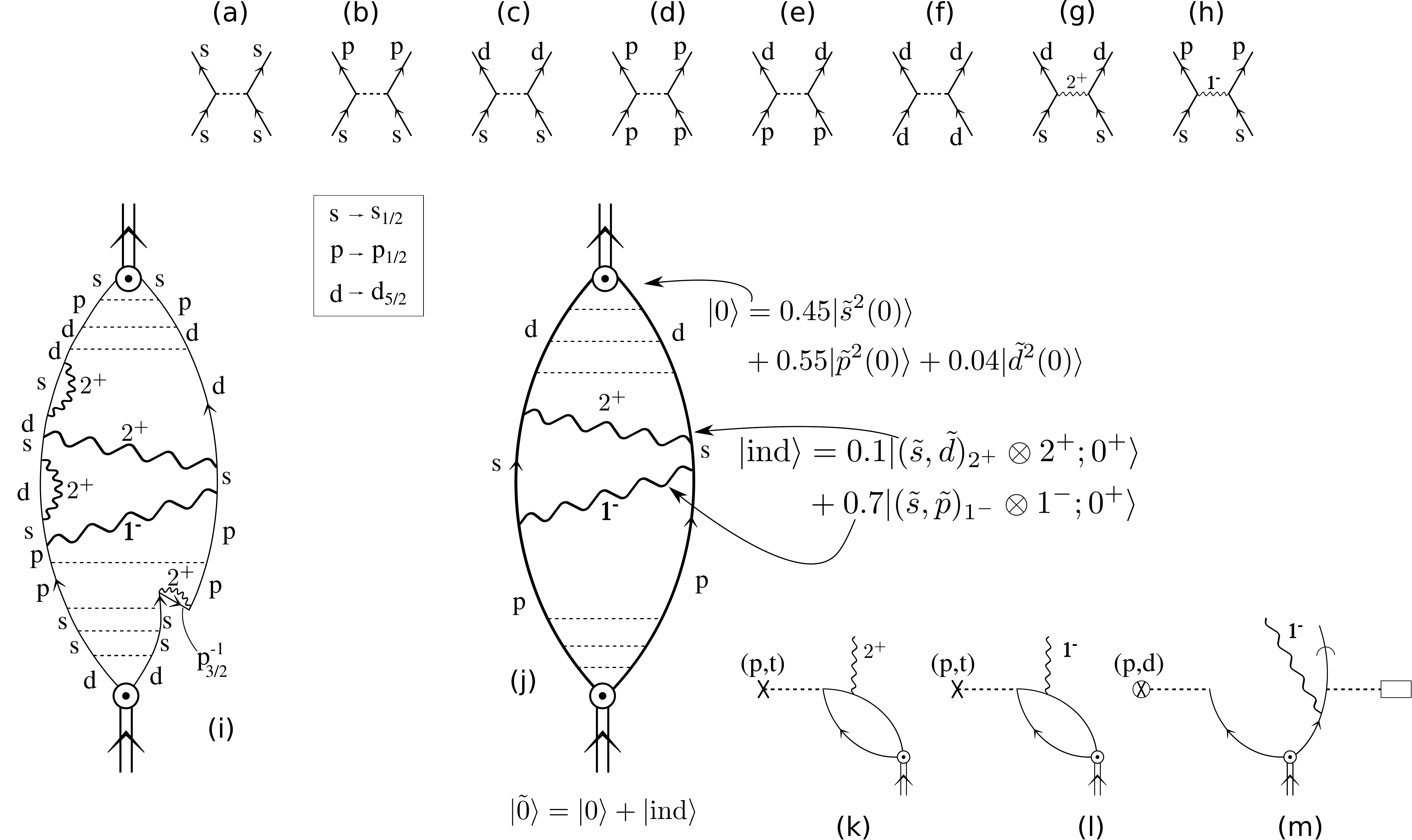}}
	\caption{Microscopic processes contributing to the structure of 
		$|{\tilde 0_{\nu}}\rangle $ (Eq. (A.1)). 
		(a)-(f): bare $^1S_0$ NN-interaction
		processes contribution to the correlation of the two-halo neutrons  of $^{11}$Li and to its binding to the core $^9$Li leading to 
		a contribution of the order of -100 keV; (g),(h) induced pairing interaction associated with the exchange of the 
		low-lying quadrupole vibration of the core and of the soft E1-mode. This last one contributes with most of the two halo neutron 
		 separation energy ($S_{2n} \approx $ 380 keV); (i) nuclear field theory diagram associated with the renormalization
		of the single-particle states, and with their binding to the $^9$Li core. Bold wavy lines describe dressed vibrational modes
		whose properties (energy and electromagnetic transition probabilities) reproduce the experimental findings (renormalised NFT);
		(j) the connection between the diagrammatic processes and the Dirac (ket)
		representation (Eqs. (A.1)-(A.3)); (k) through the invasive, irreversible $^1$H($^{11}$Li,$^9$Li($1/2^-$; 2.769 MeV))$^3$H process
		populating the $1/2^-$ member of the $(2^+\otimes p_{3/2}(\pi))_{1/2^-,...7/2^-}$ multiplet of $^9$Li, evidence for the component 
		$|(\tilde s,\tilde d)_{2^+} \times 2^+;0^+\rangle $ of $|\tilde 0_{\nu}\rangle $ was obtained \cite{Potel:10} (Fig \ref{fig:1}, see also \cite{Barranco:19b}). Similar information 
		can be gathered by recording the $\gamma-$ray associated with the decay of the quadrupole mode of $^9$Li in coincidence with the outgoing particle ($^9$Li), an experiment which remains to date of the gedanken type. However, it does not need to remain such, being a possible (and likely important) experiment, in keeping  with the fact that the $^{11}$Li beam had a rather low energy (3.3 MeV/$A$) , the flying time to the particle detector 
		after the reaction has taken place being adequate as compared to $\hbar/\Gamma_{\gamma} (\Gamma_{\gamma} \sim $eV),
		the main proviso to be made concerns the solid angle covered by the $\gamma-$detector; (l) in principle, the above 
		inverse kinematics, two-neutron pickup process can provide similar information concerning the
		(symbiotic, bootstrap) $|(\tilde s, \tilde p)_{1^-} \otimes 1^-; 0^+\rangle $ component. However, in this case, 
		the flying path is replaced by the interaction range ($\approx$ 10 fm), in keeping with the fact that the soft dipole mode is a 
		vibration which involves the neutron (halo) skin which the reaction annihilates. Because the interaction time 
		is 6-7 orders of magnitude shorter than than $\hbar/\Gamma_{\gamma}$, the process discussed  will likely remain a gedanken 
		experiment; (m) similar as before  but for the case of a single neutron pickup reaction leading to $^{10}$Li which, not being bound, 
		will loose the second neutron through coupling to the continuum (horizontal dashed line ending in an open square), the asymptotic 
		wave describing its motion being represented by a curved arrow. }
	\label{fig:5}       
\end{figure*}

The atomic nucleus provides a unique laboratory where the above concepts not only can be tested in terms of individual quantal
states, as in the case of pairing rotational bands \cite{Bes:66,Potel:13b,Hinohara:16} but also in the  limit of a single Cooper pair. 
In particular in the case of halo nuclei, much studied systems displaying large quantum fluctuations \cite{Barranco:14,Broglia:10} whose interweaving 
can be described in physical 
and numerical detail, and tested in terms of absolute differential  cross sections.

\begin{figure*}
\resizebox{0.9\textwidth}{!}{%
  \includegraphics{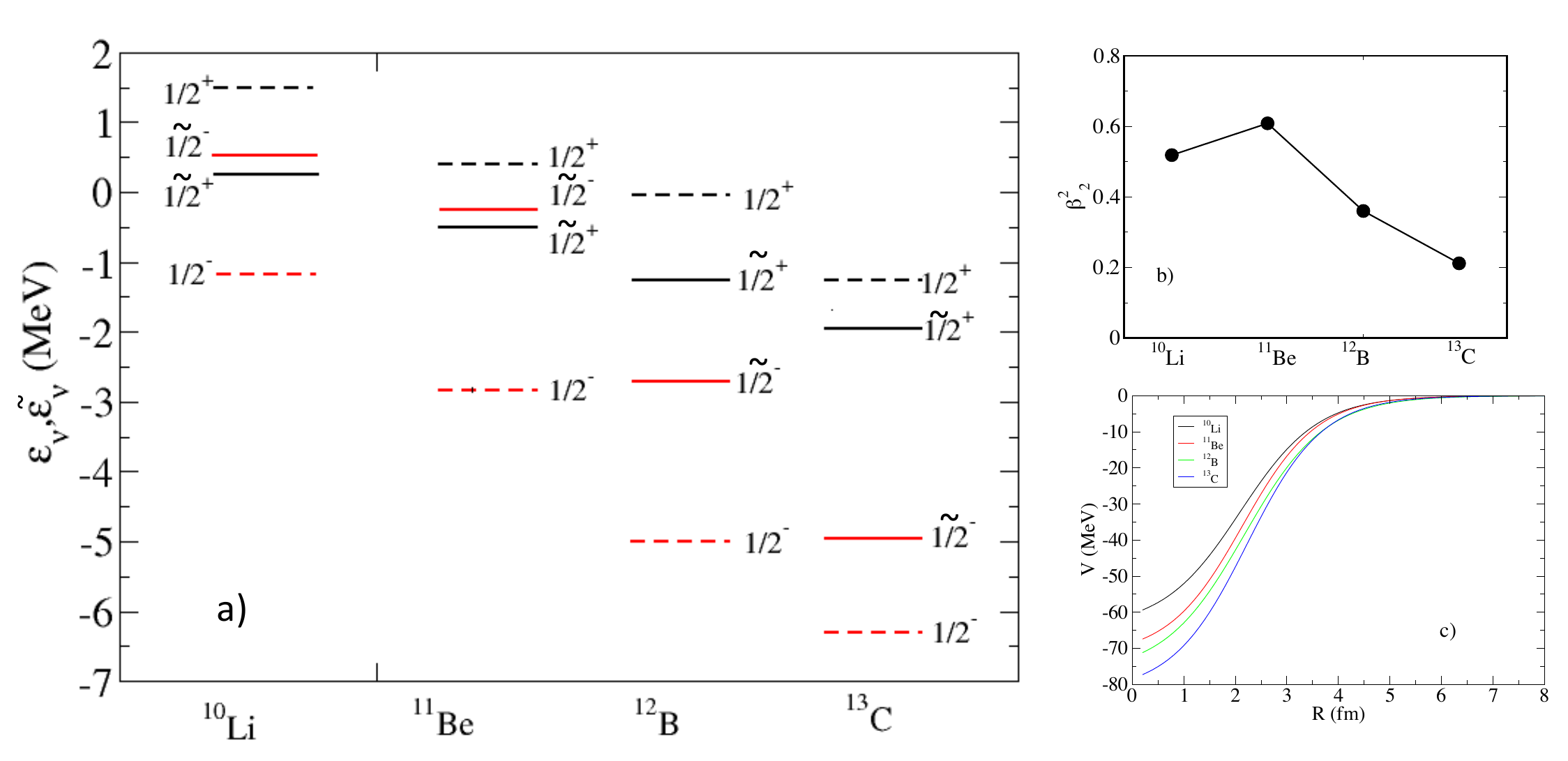}
}
\caption{(color online)
Bare ($\epsilon_{\nu}$, horizontal dashed) and renormalised
($\tilde \epsilon_{\nu}$, horizontal continuous  bold face) single-particle energies associated with the $1/2^+$  (black lines) 
and $1/2^-$ (red lines)  states of the 
$N=7$ isotones.  (b) The square of the dynamical deformation $\beta_2$ associated with the lowest quadrupole vibration of these nuclei
are indicated with a solid dot. (c) The bare mean field potentials.
}
\label{fig:6}       
\end{figure*}

Within this scenario, a particularly  attractive system is $^{12}_4$Be$_8$, with two neutrons outside the N = 6 closed shell\footnote{In this system, it is the first excited $0^+$ state which can be viewed as a neutron halo Cooper pair, similar to the di-neutron component entering   the $^{11}$Li ground state (see footnote 3).}.
This non-(Meyer-Jensen) magic number results from the phenomenon of  parity inversion. Namely, from the quantal  phase transition 
observed in $^{11}$Be and resulting from the crossing  of the $1p_{1/2}$ and the $2s_{1/2}$ levels  and leading to the  $1/2^+$
ground state ($S_n = 0.5$ MeV) and to the first excited $1/2^-$ state ($E_x = 0.32$ MeV). 
At the basis of these phenomenon one finds the self energy process controlled essentially by the coupling 
of the  $1p_{1/2}$ and the $2s_{1/2}$ states with the  low-lying quadrupole  vibration of the core ($\hbar \omega_2 = 3.368 $ MeV ($\beta_2 \approx 0.9)$
\cite{Barranco:17}).  This implies a change from a regime in which static mean field effects dominate over quantal fluctuations, to another one in which the situation is essentially reversed. Saying it differently with the help of the definitions  
\begin{equation}
U(r)= \int d^3 r' \rho(r') v(|\vec r - \vec r'|),
\end{equation}
and
\begin{equation}
\delta U (r) = \int  d^3 r' \delta \rho(r') v(|\vec r -\vec r'|),
\end{equation}
from a situation in which $\bar U \gg\delta \bar U$, 
 to one in which $\delta \bar U \geq  \bar U$.  The function $U(r)$ is the Hartree potential while $\delta U$ describes the quantal fluctuations of $U$
 associated with vibrations of the system, in particular of the nuclear surface, while $\bar U = \langle \phi_{\nu_F}|U|\phi_{\nu_F}\rangle  $ and similarly 
$ \delta \bar U = \langle \phi_{\nu_F}|\delta U | \phi_{\nu_F} \rangle $, $\phi_{\nu_F}$ describing the single-particle  motion of a nucleon at the Fermi energy. 
 

 Writing the Hamiltonian as $H= H_0 + \beta_2 H' ([H_0,H']=0)$, where $H_0 =  T+U$ and 
 with 
 $H' = \left(\frac{1}{16\pi} \right)^{1/2} \langle R_0 \partial U/\partial r\rangle $ (see \cite{Bohr:75}, Eq. (6-68)) one observes that, starting  from the N=7 isotope $^{13}_6$C 
 displaying  the standard $1p_{1/2}-2s_{1/2}$ sequence, and removing one proton at a time, 
 the (quadrupole) particle-vibration coupling  strength 
 changes leading eventually to a crossing of the $1/2^-,1/2^+$ levels, this last becoming ground state for $^{11}$Be (Fig. \ref{fig:6}). 
 Parity inversion also found 
 in $^{10}$Li (\cite{Barranco:19} and refs. therein) \footnote{While the bare energies of the single-particle states are not 
 observable, the $B(E2)$ transitions  probabilities ($\beta_2^2)$ of the low-lying collective states are so. The microscopic 
 mechanism at the basis of parity inversion in $^{11}$Be and $^{10}$Li is parallel to that at the basis of the Lamb shift in the hydrogen atom (H). 
 Namely self energy and Pauli principle (see \cite{Weinberg:96}, Fig. 14.2,  \cite{Barranco:17}  Figs. 2(I)(a) and 2(II)(b)). In (renormalised) QED the bare electron mass is the (divergent) quantity one has to subtract  so that  the renormalised energy splitting between the $^2S_{1/2}$ and $^2P_{1/2}$ levels of the 
 H atom best reproduces  the experimental findings (1057.845(9) MHz (Exp.), 1057.865 MHz). In (renormalised) nuclear field theory, the (four) parameters of the bare Saxon-Woods potential plus $k-$mass are adjusted so as the renormalized single-particle energies best
 reproduce the observed quantities (see Fig. \ref{fig:6}). Within this context it is of notice the remarkable  prediction of parity inversion  made in
 \cite{Talmi:60}, and the detailed, profound analysis of the phenomenon presented in \cite{Sagawa:93}.}.

 Being these systems at zero temperature, we are in presence of a quantal  phase transition. Phase transition which induces   the tumbling chain 
 parity-inversion $\to$ presence of (dynamical) dipole moment in the ground state of \Liel. 
 More accurately, the emergence of a soft E1-mode 
 carrying about 6\% of the TRK sum rule.  
 Tumbling which, in the case of \Liel, becomes a bootstrap circular chain. In fact, the soft E1-mode of \Liel,
 which consists in a vibrational motion of the halo neutrons against the protons and neutrons of the core which vibrate in phase, acts as 
 the glue of the two halo neutrons to the core $^{9}$Li and thus to the very existence of the halo. 
 
 Given the success of BCS theory in the case  of superconductivity, it was natural to ask whether a similar mechanism  may also work for the 
 fermionic superfluid $^3_2$He$_1$. In this case the attraction between the atoms must be an intrinsic property of $^3$He. The main feature of the inter-atomic potential is the strong repulsive component at short distances, and the weak Van der Waals  attraction at medium and large distances. To avoid repulsion the $^3$He atoms have to form
Cooper pairs  in a state of  relative angular momentum $L $ different from zero, thus being kept away by the centrifugal barrier. It turned out that $L=1$ and  $S=1$
\cite{Vollhardt:90}.

Early experimental data indicated that the entities forming the Cooper pairs  were  not the bare $^3$He atoms,  but strongly dressed fermions, 
displaying effective masses about six times larger than the bare atomic mass 
\footnote{In the case of $^{11}$Li, the mass increase from the bare (4/2 MeV$^{-1}$, 4 being the summed degeneracy of the $p_{1/2}$ and
$s_{1/2}$ states, 2 MeV their energy separation) to the dressed one (4/0.4 MeV$^{-1}$) is of a factor of 5, in keeping with the fact that the density 
of levels around the Fermi energy is inversely proportional to the effective mass.}. 
It is thus not surprising that the bare  atomic potential bears little resemblance to the effective potential acting  between the dressed fermions 
(quasiparticles).
The most important consequence of
 the onset of Cooper pairing in liquid $^3$He  is the ability to amplify ultra weak effects, in particular those  of the electromagnetic interaction between the nuclear dipole moments. Even at the distance of closest approach of two $^3$He atoms (the hard-core radius R$\approx 2 \AA$) this interaction is only of
  the order of $10^{-7}K$, orders of magnitude smaller than the critical temperature ($\approx 10^{-3}$K). The amplification 
  arises due to the fact that the system mediating the induced interaction (the quasiparticles) is the very same system between which the interaction operates
  (bootstrap) \footnote{As stated above, in $^{11}$Li the system mediating  the interaction between the two halo neutrons (the halo field) is the very same system
  through which  the induced  interaction operates (exchange of soft E1-mode between the halo neutrons).}.
  
  It is of notice that while the direct interaction between the nuclear dipole moments of two $^3$He atoms   is invariant under simultaneous rotation of $\vec L$ and $\vec S$,
  it is not invariant under rotation of either alone. Consequently, if the ``strong''
  (Van der Waals, etc.) force in the problem forces all pairs to have e.g. a fixed and  identical relation of
  their orbital orientation to their spin,  the condensate will  display a spontaneously broken spin-orbit symmetry  (SBSOS) \cite{Leggett:06}.  That is, the
  symmetries  with respect to rotation in spin and orbital space are spontaneously broken, leading to a highly degenerate  ground state. In this situation, a tiny dipole-dipole 
  interaction, which implies a spin-orbit coupling, is able to lift the degeneracy, by choosing that particular relative orientation of $\vec L$ and $\vec S$
  for which the dipolar energy is minimal and in this case of the order of $N_p \times 10^{-7}$K ($N_p \approx 10^{22}$, number of Cooper pairs). 
  Thereby the interaction between the nuclear dipole moments acquires macroscopic relevance \cite{Anderson:66}.
  
  Central  nuclear forces, in which the line of action passes through the pair of nucleons although  their magnitude may depend on the relative orientation 
  of their spins  conserve $\vec S $ and $\vec L$ separately, as well as the total angular momentum $\vec J$ and parity $\pi$.
  The tensor force $V(r)= V_T (r) \{3(\vec \sigma_1 \cdot \hat r)(\vec \sigma_2 \cdot \hat r) - \vec \sigma_1 \cdot \vec \sigma_2\}$,
  where $\sigma_1$ and $\sigma_2$ are the Pauli matrices of the  two nucleons, $\hat r$ is the unit vector 
  parallel to the line joining them and $V_T(r)$ is some scalar function of their separation,  is non-central and thus mixes states  with the same
  $\vec J$ but different  $\vec L$ and $\vec S$ \cite{Brink:55b}.

While the tensor force  is rather strong its influence on the properties of the nucleus still remains an open problem. 
This is to some extent due to the fact that its effects are mixed  
 up with those of the spin-orbit force $V_{so} (r) \vec L \cdot  \vec S$.
 
 To which extent the tumbling (bootstrap) chain qua\-dru\-pole (dynamical) surface distortion $\to$ parity  inversion (level crossing) $\to$ dipole 
 (dynamical) distortion  $\to$ (incipient) gauge invariance distortion at the single Cooper pair level found in the NFT description of $^{11}$Li can be related to the tensor force  
 \cite{Otsuka:05,Myo:07} and, arguably, to some kind of SBSOS, is an open and challenging question.
   

\begin{thebibliography}{10}
	
	\bibitem{Bortignon:78}
	{Bortignon, P. F.}, R.~A. Broglia, and D.~R. B{\`{e}}s.
	\newblock On the convergence of nuclear field theory perturbation expansion for
	strongly anharmonic systems.
	\newblock {\em Phys. Lett. B}, 76:153, 1978.
	
	\bibitem{Bortignon:77b}
	P.~F. Bortignon.
	\newblock Nuclear field theory of two-phonon states.
	\newblock In A.~Bohr and R.~A. Broglia, editors, {\em International School of
		Physics ``Enrico Fermi'' Course LXIX, Elementary Modes of Excitation in
		Nuclei}, page 519, Amsterdam, 1977. North Holland.
	
	\bibitem{Repko:13}
	A.~Repko, P.~G. Reinhard, V.~O. Nesterenko, and J.~Kvasil.
	\newblock {Toroidal nature of the low--energy E1 mode}.
	\newblock {\em Phys. Rev. C}, 87:04305, 2013.
	
	\bibitem{Ryezayeva:02}
	N.~Ryezayeva, T.~Hartmann, Y.~Kalmykov, H.~Lenske, P.~von Neumann-Cosel, V.~Yu.
	Ponomarev, A.~Richter, A.~Shevchenko, S.~Volz, and J.~Wambach.
	\newblock Nature of low-energy dipole strength in nuclei: The case of a
	resonance at particle threshold in $^{\mathrm{208}}\mathrm{P}\mathrm{b}$.
	\newblock {\em Phys. Rev. Lett.}, 89:272502, 2002.
	
	\bibitem{Broglia:71}
	R.~A. Broglia, C.~Riedel, and T.~Udagawa.
	\newblock Coherence properties of two-neutron transfer reactions and their
	relation to inelastic scattering.
	\newblock {\em Nuclear Physics A}, 169:225, 1971.
	
	\bibitem{Cooper:56}
	L.~N. Cooper.
	\newblock {Bound Electron Pairs in a Degenerate Fermi Gas}.
	\newblock {\em Phys. Rev.}, 104:1189, 1956.
	
	\bibitem{Barranco:01}
	{Barranco, F.}, P.~F. Bortignon, R.~A. Broglia, G.~Col{\`{o}}, and E.~Vigezzi.
	\newblock The halo of the exotic nucleus $^{11}${Li}: a single {C}ooper pair.
	\newblock {\em Europ. Phys. J. A}, 11:385, 2001.
	
	\bibitem{Zhukov:93}
	M.V. Zhukov, B.V. Danilin, D.V. Fedorov, J.M. Bang, I.J. Thompson, and J.S.
	Vaagen.
	\newblock {Bound state properties of Borromean halo nuclei: $^{6}$He and
		$^{11}$Li}.
	\newblock {\em Physics Reports}, 231(4):151 -- 199, 1993.
	
	\bibitem{Josephson:62}
	B.~D. Josephson.
	\newblock Possible new effects in superconductive tunnelling.
	\newblock {\em Phys. Lett.}, 1:251, 1962.
	
	\bibitem{Tanihata:08}
	{Tanihata, I.}, M.~Alcorta, D.~Bandyopadhyay, R.~Bieri, L.~Buchmann, B.~Davids,
	N.~Galinski, D.~Howell, W.~Mills, S.~Mythili, R.~Openshaw, E.~Padilla-Rodal,
	G.~Ruprecht, G.~Sheffer, A.~C. Shotter, M.~Trinczek, P.~Walden, H.~Savajols,
	T.~Roger, M.~Caamano, W.~Mittig, P.~Roussel-Chomaz, R.~Kanungo, A.~Gallant,
	M.~Notani, G.~Savard, and I.~J. Thompson.
	\newblock Measurement of the two-halo neutron transfer reaction
	{$^1$H($^{11}$Li,$^{9}$Li)$^3$H} at {3A} {MeV}.
	\newblock {\em Phys. Rev. Lett.}, 100:192502, 2008.
	
	\bibitem{Potel:10}
	G.~Potel, F.~Barranco, E.~Vigezzi, and R.~A. Broglia.
	\newblock {Evidence for phonon mediated pairing interaction in the halo of the
		nucleus $^{11}$Li}.
	\newblock {\em Phys. Rev. Lett.}, 105:172502, 2010.
	
	\bibitem{Kanungo:15}
	R.~Kanungo, A.~Sanetullaev, J.~Tanaka, S.~Ishimoto, G.~Hagen, T.~Myo,
	T.~Suzuki, C.~Andreoiu, P.~Bender, A.~A. Chen, B.~Davids, J.~Fallis, J.~P.
	Fortin, N.~Galinski, A.~T. Gallant, P.~E. Garrett, G.~Hackman, B.~Hadinia,
	G.~Jansen, M.~Keefe, R.~Kr\"ucken, J.~Lighthall, E.~McNeice, D.~Miller,
	T.~Otsuka, J.~Purcell, J.~S. Randhawa, T.~Roger, A.~Rojas, H.~Savajols,
	A.~Shotter, I.~Tanihata, I.~J. Thompson, C.~Unsworth, P.~Voss, and Z.~Wang.
	\newblock Evidence of soft dipole resonance in $^{11}\mathrm{Li}$ with
	isoscalar character.
	\newblock {\em Phys. Rev. Lett.}, 114:192502, 2015.
	
	\bibitem{Nakamura:06}
	{T. Nakamura {et al.}}
	\newblock {Observation of Strong Low-Lying E1 Strength in the Two-Neutron Halo
		Nucleus $^{11}$Li}.
	\newblock {\em Phys. Rev. Lett.}, 96:252502, 2006.
	
	\bibitem{Avogadro:07}
	P.~Avogadro, F.~Barranco, R.~A. Broglia, and E.~Vigezzi.
	\newblock Quantum calculation of vortices in the inner crust of neutron stars.
	\newblock {\em Phys. Rev. C}, 75:012805, 2007.
	
	\bibitem{Avogadro:08}
	P.~Avogadro, F.~Barranco, R.A. Broglia, and E.~Vigezzi.
	\newblock {Vortex--nucleus interaction in the inner crust of neutron stars}.
	\newblock {\em Nuclear Physics A}, 811(3):378, 2008.
	
	\bibitem{Broglia:19}
	R.A. Broglia, F.~Barranco, A.~Idini, G.~Potel, and E.~Vigezzi.
	\newblock Pygmy resonances: what's in a name?
	\newblock {\em arXiv:18096.09409; Phys. Scr., in press}, 2019.
	
	\bibitem{Lenske:01}
	H.~Lenske, F.~Hofmann, and C.M. Keil.
	\newblock Probing isospin dynamics in halo nuclei.
	\newblock {\em Progress in Particle and Nuclear Physics}, 46(1):187 -- 196,
	2001.
	
	\bibitem{Orrigo:09}
	{Orrigo, S. E. A.} and H.~Lenske.
	\newblock {Pairing resonances and the continuum spectroscopy of $^{10}$Li}.
	\newblock {\em Phys. Lett. B}, 677:214, 2009.
	
	\bibitem{Barranco:19}
	F.~Barranco, G.~Potel, E.~Vigezzi, and R.~A. Broglia.
	\newblock {$d(^9$Li,$p$), specific probe of $^{10}$Li, paradigm of
		parity--inverted, soft--dipole isotones with one neutron outside the $N = 6$
		closed shell}.
	\newblock {\em arXiv:1812.01761}.
	
	\bibitem{Broglia:16}
	R.~A. Broglia, P.~F. Bortignon, F.~Barranco, E.~Vigezzi, A.~Idini, and
	G.~Potel.
	\newblock {Unified description of structure and reactions: implementing the
		Nuclear Field Theory program}.
	\newblock {\em Phys. Scr.}, 91:063012, 2016.
	
	\bibitem{Bes:63}
	D.~R. B{\`{e}}s.
	\newblock Beta--vibrations in even nuclei.
	\newblock {\em Nuclear Physics}, 49:544 -- 565, 1963.
	
	\bibitem{Aprahamian:18}
	A.~Aprahamian, R.~C. de~Haan, S.~R. Lesher, C.~Casarella, A.~Stratman, H.~G.
	B\"orner, H.~Lehmann, M.~Jentschel, and A.~M. Bruce.
	\newblock Lifetime measurements in $^{156}\mathrm{Gd}$.
	\newblock {\em Phys. Rev. C}, 98:034303, 2018.
	
	\bibitem{Sharpey:19}
	{Sharpey-Schafer, J. F.}, {Bark, R. A.}, {Bvumbi, S. P.}, {Dinoko, T. R. S.},
	and {Majola, S. N. T.}
	\newblock "stiff" deformed nuclei, configuration dependent pairing and the
	$\beta$ and $\gamma$ degrees of freedom.
	\newblock {\em Eur. Phys. J. A}, 55(2):15, 2019.
	
	\bibitem{Maher:70}
	J.~V. Maher, J.~R. Erskine, A.~M. Friedman, J.~P. Schiffer, and R.~H. Siemssen.
	\newblock Unexpected strong pair correlations in excited ${0}^{+}$ states of
	actinide nuclei.
	\newblock {\em Phys. Rev. Lett.}, 25:302--306, 1970.
	
	\bibitem{Casten:72}
	R.F. Casten, E.R. Flynn, J.D. Garrett, O.~Hansen, T.J. Mulligan, D.R.
	B{\`{e}}s, R.A. Broglia, and B.~Nilsson.
	\newblock {Search for $(t, p)$ transitions to excited $0^+$ states in the
		actinide region}.
	\newblock {\em Physics Letters B}, 40:333, 1972.
	
	\bibitem{Ragnarsson:76}
	I.~Ragnarsson and R.~A. Broglia.
	\newblock Pairing isomers.
	\newblock {\em Nuclear Physics A}, 263:315, 1976.
	
	\bibitem{Mukerjee:89}
	Madhusree Mukerjee and Yoichiro Nambu.
	\newblock {BCS and IBM}.
	\newblock {\em Annals of Physics}, 191(1):143 -- 162, 1989.
	
	\bibitem{Potel:13b}
	{Potel, G.}, A.~Idini, F.~Barranco, E.~Vigezzi, and R.~A. Broglia.
	\newblock {Quantitative study of coherent pairing modes with two--neutron
		transfer: Sn isotopes}.
	\newblock {\em {Phys. Rev. C}}, 87:054321, 2013.
	
	\bibitem{Barranco:88}
	{Barranco, F.}, R.~A. Broglia, and G.~F. Bertsch.
	\newblock Exotic radioactivity as a superfluid tunneling phenomenon.
	\newblock {\em Phys. Rev. Lett.}, 60:507, 1988.
	
	\bibitem{Barranco:90}
	{Barranco, F.}, G.F. Bertsch, R.A. Broglia, and E.~Vigezzi.
	\newblock Large-amplitude motion in superfluid fermi droplets.
	\newblock {\em Nuclear Physics A}, 512:253, 1990.
	
	\bibitem{Brink:05}
	{Brink, D.} and R.~A. Broglia.
	\newblock {\em Nuclear Superfluidity}.
	\newblock Cambridge University Press, Cambridge, 2005.
	
	\bibitem{Potel:13}
	{Potel, G.}, A.~Idini, F.~Barranco, E.~Vigezzi, and R.~A. Broglia.
	\newblock Cooper pair transfer in nuclei.
	\newblock {\em {Rep. Prog. Phys.}}, 76:106301, 2013.
	
	\bibitem{Broglia:04a}
	R.~A. Broglia and A.~Winther.
	\newblock {\em Heavy Ion Reactions}.
	\newblock Westview Press, Boulder, CO., 2004.
	
	\bibitem{Potel:x}
	G.~Potel.
	\newblock {\textsc{cooper}, code for two-nucleon transfer reactions}.
	\newblock {\em (unpublished)}.
	
	\bibitem{Bardeen:57a}
	J.~Bardeen, L.~N. Cooper, and J.~R. Schrieffer.
	\newblock Microscopic theory of superconductivity.
	\newblock {\em Physical Review}, 106:162, 1957.
	
	\bibitem{Bardeen:57b}
	J.~Bardeen, L.~N. Cooper, and J.~R. Schrieffer.
	\newblock Theory of superconductivity.
	\newblock {\em Physical Review}, 108:1175, 1957.
	
	\bibitem{Nambu:91}
	Y.~Nambu.
	\newblock Dynamical symmetry breaking.
	\newblock In T.~Eguchi and K.~Nishigama, editors, {\em Broken Symmetry,
		Selected papers of Y. Nambu (1995)}, page 436. World Scientific, Singapore,
	1991.
	
	\bibitem{Barranco:19b}
	F.~Barranco, G.~Potel, E.~Vigezzi, and R.~A. Broglia.
	\newblock {Radioactive beams and inverse kinematics: probing the quantal
		texture of the nuclear vacuum}.
	\newblock {\em arXiv:1904.02786}, 2019.
	
	\bibitem{Bes:66}
	{B{\`{e}}s, D. R.} and R.~A. Broglia.
	\newblock Pairing vibrations.
	\newblock {\em Nucl. Phys.}, 80:289, 1966.
	
	\bibitem{Hinohara:16}
	Nobuo Hinohara and Witold Nazarewicz.
	\newblock {Pairing Nambu-Goldstone Modes within Nuclear Density Functional
		Theory}.
	\newblock {\em Phys. Rev. Lett.}, 116:152502, 2016.
	
	\bibitem{Barranco:14}
	F.~Barranco, R.~A. Broglia, G.~Potel, and E.~Vigezzi.
	\newblock Core polarization and neutron halos.
	\newblock {\em Journal of Physics: Conference Series}, 527:012005, 2014.
	
	\bibitem{Broglia:10}
	R.~A. Broglia, G.~Potel, F.~Barranco, and E.~Vigezzi.
	\newblock Difference between stable and exotic nuclei: medium polarization
	effects.
	\newblock {\em J. Phys. G}, 37:064022, 2010.
	
	\bibitem{Barranco:17}
	F.~Barranco, G.~Potel, R.~A. Broglia, and E.~Vigezzi.
	\newblock {Structure and reactions of $^{11}$Be: many--body basis for
		single--neutron halo}.
	\newblock {\em Phys. Rev. Lett.}, 119:082501, 2017.
	
	\bibitem{Bohr:75}
	{Bohr, A.} and B.~R. Mottelson.
	\newblock {\em Nuclear Structure, Vol.II}.
	\newblock Benjamin, New York, 1975.
	
	\bibitem{Weinberg:96}
	S.~Weinberg.
	\newblock {\em The Quantum Theory of Fields}, volume~1.
	\newblock Cambridge University Press, Cambridge, 1996.
	
	\bibitem{Talmi:60}
	I.~Talmi and I.~Unna.
	\newblock {Order of Levels in the Shell Model and Spin of
		${\mathrm{Be}}^{11}$}.
	\newblock {\em Phys. Rev. Lett.}, 4:469, 1960.
	
	\bibitem{Sagawa:93}
	H.~Sagawa, B.~A. Brown, and H.~Esbensen.
	\newblock {Parity inversion in the N=7 isotones and the pairing blocking
		effect}.
	\newblock {\em Phys. Lett. B}, 309:1, 1993.
	
	\bibitem{Vollhardt:90}
	D.~Vollhardt and P.~W{\"{o}}lfle.
	\newblock {\em The superfluid phases of Helium 3}.
	\newblock Taylor and Francis, London, 1990.
	
	\bibitem{Leggett:06}
	A.J. Leggett.
	\newblock {\em Quantum Liquids}.
	\newblock Oxford University Press, Oxford, 2006.
	
	\bibitem{Anderson:66}
	P.~W. Anderson.
	\newblock Considerations on the flow of superfluid helium.
	\newblock {\em Rev. Mod. Phys.}, 38:298--310, 1966.
	
	\bibitem{Brink:55b}
	D.~M. Brink.
	\newblock {\em {Nuclear Forces}}.
	\newblock Pergamon Press, Oxford, 1955.
	
	\bibitem{Otsuka:05}
	Takaharu Otsuka, Toshio Suzuki, Rintaro Fujimoto, Hubert Grawe, and Yoshinori
	Akaishi.
	\newblock Evolution of nuclear shells due to the tensor force.
	\newblock {\em Phys. Rev. Lett.}, 95:232502, 2005.
	
	\bibitem{Myo:07}
	Takayuki Myo, Kiyoshi Kat\ifmmode~\bar{o}\else \={o}\fi{}, Hiroshi Toki, and
	Kiyomi Ikeda.
	\newblock Roles of tensor and pairing correlations on halo formation in
	$^{11}\mathrm{Li}$.
	\newblock {\em Phys. Rev. C}, 76:024305, 2007.
	
\end{thebibliography}

\end{document}